\documentclass[reprint,aps,prd,twocolumn,superscriptaddress,nofootinbib,floatfix]{revtex4-2}

\usepackage[utf8]{inputenc}
\usepackage{amsmath, amssymb, amsfonts, bm}
\usepackage{graphicx}
\usepackage[colorlinks=true, linkcolor=blue, citecolor=blue, urlcolor=blue]{hyperref}
\usepackage{microtype}
\usepackage{orcidlink}
\DeclareUnicodeCharacter{2299}{\odot}

\begin{document}

\title{Search for Eccentric Neutron Star--Black Hole Binaries in Advanced LIGO's Fourth Observing Run using Effective-One-Body Waveforms}

\author{{Aleyna Akyüz}\orcidlink{0000-0001-7248-951X}}
\affiliation{Department of Physics, Syracuse University, Syracuse, New York 13244, USA}

\author{{Keisi Kacanja}\orcidlink{0009-0004-9167-7769}}
\affiliation{Department of Physics, Syracuse University, Syracuse, New York 13244, USA}

\author{{Rahul Dhurkunde}\orcidlink{0000-0002-5077-8916}}
\affiliation{Institute of Cosmology and Gravitation, University of Portsmouth, Portsmouth, PO1 3FX, U.K.}

\author{Alexander H. Nitz\orcidlink{0000-0002-1850-4587}}
\email{ahnitz@syr.edu}
\affiliation{Department of Physics, Syracuse University, Syracuse, New York 13244, USA}

\date{\today}

\begin{abstract}
The observation of orbital eccentricity in compact binary coalescences would provide a direct signature of dynamical formation environments. Previous matched-filter searches for eccentric neutron star-black hole (NSBH) binaries have been severely limited by computational constraints, necessitating the use of simplified waveform models that capture secular dephasing but omit apsidal precession and inter-orbit harmonics. Including these physical features increases the sensitive search volume for systems with $0.2 \le e_{20} \le 0.28$ by $\sim 25\%$ to $30\%$. In this work, we present a rapid matched-filter search methodology that makes the direct use of time-domain effective-one-body waveforms computationally feasible across a wide parameter space. By multibanding and high-frequency ratio-filtering, our pipeline achieves a $\sim 20\times$ reduction in computational cost on high-throughput clusters. This allows a bank of 88.6 million effective-one-body templates, fifteen times larger than prior eccentric analyses, to search the first two observing epochs of the fourth observing run (O4a and O4b) of Advanced LIGO. We report the results of this search across both coincident and non-coincident observing time, totaling 401~days (comprising 238~days of coincident H1L1 data and 163~days of single-detector data). We recover the known low-mass mergers GW230518\_125908 and GW230529\_181500 (the latter identified during single-detector observing time) with parameters consistent with zero eccentricity. Finding no new gravitational-wave events with orbital eccentricity ($e_{20} \ge 0.05$), we place $90\%$ credible upper limits on the local merger rate density of eccentric NSBH systems across the search space of $\mathcal{R}_{90\%} \le 7.5\text{--}57\ \mathrm{Gpc}^{-3}\ \mathrm{yr}^{-1}$, up to a factor of $\sim 4$ improvement over previous observational bounds. At the parameters of GW200105\_162426, this non-detection establishes an upper limit of $\mathcal{R}_{90\%} \le 12.1\ \mathrm{Gpc}^{-3}\ \mathrm{yr}^{-1}$. Alternatively, if GW200105\_162426 is treated as an eccentric detection, the lack of subsequent detections in O4 updates the inferred local merger rate density of this population to $\mathcal{R} = 6.6^{+6.3}_{-3.8}\ \mathrm{Gpc}^{-3}\ \mathrm{yr}^{-1}$ ($90\%$ credible upper bound of $15.3\ \mathrm{Gpc}^{-3}\ \mathrm{yr}^{-1}$).
\end{abstract}

\maketitle


\textit{Introduction.}---Gravitational-wave astronomy has now become a routine tool for how we observe the Universe.
The second-generation of gravitational-wave observatories, which began operation back in 2015 and includes Advanced LIGO~\cite{LIGOScientific:2014pky} and Advanced Virgo~\cite{VIRGO:2014yos} now routinely observe the coalescence of compact binaries. To date, hundreds of binary black hole (BBH)~\cite{LIGOScientific:2025slb,LIGOScientific:2026wfs,Nitz:2021zwj}, a handful of neutron-star black-hole (NSBH)~\cite{LIGOScientific:2021qlt,LIGOScientific:2024elc}, and two binary neutron star (BNS)~\cite{LIGOScientific:2017vwq,LIGOScientific:2020aai} sources have been observed. Individually, these observations have been useful for probing beyond-GR theories~\cite{LIGOScientific:2026oim}, the nuclear equation-of-state~\cite{LIGOScientific:2018cki}, and have helped establish the neutron star merger origin of Gamma Ray Bursts (GRBs)~\cite{LIGOScientific:2017zic,LIGOScientific:2017ync}. 

The large population of observed binary black hole mergers is starting to help shape our understanding of the history and formation of compact binaries~\cite{LIGOScientific:2026ctl,Mapelli:2020vfa}. There are numerous 
proposals for how these binaries form and evolve~\cite{Mapelli:2020vfa}; broadly speaking they can be grouped
into various isolated~\cite{Marchant:2023wno,Belczynski:2016obo} and dynamical~\cite{Rodriguez:2016vmx,Samsing:2017xmd,Rodriguez:2018rmd} formation scenarios. While it is currently unclear how many distinct formation channels are contributing to the observed population~\cite{LIGOScientific:2026ctl}, an emerging consensus~\cite{LIGOScientific:2026ctl,Stegmann:2025zkb} is that dynamic formation is a contributing channel, especially for the more massive of the observed population. There are several lines of evidence for this scenario; these include the observation of binaries within the supernovae pair-instability mass gap~\cite{LIGOScientific:2020iuh,LIGOScientific:2020ufj}, the observation of in-plane spins~\cite{KAGRA:2021duu}, potential precessing binaries~\cite{Hannam:2021pit, LIGOScientific:2025brd}, and correlations between mass and spin~\cite{LIGOScientific:2026ctl}. There is also the possibility that some sources have non-zero orbital eccentricity~\cite{Gupte:2024jfe,Romero-Shaw:2025vbc,Gamba:2021gap}; although other explanations are possible~\cite{Romero-Shaw:2024klf,Mishra:2025dpa}, this would be consistent with a dynamical environment~\cite{Samsing:2017xmd,Zevin:2018kzq}. Despite this, these scenarios do not necessarily explain all the features of the observed population at the same time~\cite{LIGOScientific:2026ctl}.

While the picture for BBH formation is starting to come into a coarse focus, the evidence for formation of BNS and NSBH binaries primarily comes from non-GW observations~\cite{Tauris:2017omb, Chattaraj:2025rig, Grichener:2026nph}. These include observations of galactic and extragalactic sources electromagnetically. It is clear that the galactic fields provide one source for neutron star binaries, though it is not yet clear if the merging population we observe with gravitational waves is identical to those we observe with EM observatories. We have weaker evidence for the source of neutron-star black hole binaries, though isolated formation in the field has been the prevailing idea~\cite{Broekgaarden:2021iew,Broekgaarden:2021hlu,Chattopadhyay:2022cnp}. This expectation has been challenged by evidence for eccentricity in a NSBH binary GW200105\_162426~\cite{Morras:2025xfu,Planas:2025plq,Kacanja:2025kpr,Jan:2025fps}. Although the evidence is inconclusive~\cite{Clarke:2026cuw,Kacanja:2025kpr}, if the observation of eccentricity in the NSBH population is confirmed this would require either that the binary has recently formed through a dynamical mechanism or is in a triple system where Kozai-Lidov resonances could drive up the eccentricity~\cite{Fragione:2019zhm,Romero-Shaw:2025otx}. Eccentricity is such a strong marker for a dynamical environment because it is efficiently burned off through gravitational radiation \cite{Peters:1964zz}; this is the reason that most binaries are expected to have negligible eccentricity by the time they are observed by ground-based gravitational-wave observatories.

There have been a few systematic searches for compact binaries with non-negligible eccentricity; these have targeted BNS~\cite{Nitz:2019spj,Dhurkunde:2023qoe}, NSBH~\cite{Dhurkunde:2023qoe}, BBH~\cite{LIGOScientific:2019dag,LIGOScientific:2023lpe,Wang:2025yac} sources, and a targeted eccentric search on a limited segment of O3 data  to detect GW200105\_162426 \cite{Phukon:2025cky}. These searches include both loosely modelled analyses~\cite{KAGRA:2021tnv,KAGRA:2021bhs,LIGOScientific:2023lpe}, which are most sensitive to short-duration (the highest BBH) signals, and matched-filter analyses that filter the data with a model of the gravitational wave signal. Searches that do not explicitly account for eccentricity also do retain some sensitivity to low-eccentricity sources ($e_{20}\sim 0.1$) ~\cite{Ramos-Buades:2020eju,Gadre:2024ndy}. In fact, all signals that currently have support for eccentricity were initially identified by searches that did not model eccentricity at all. However, as the eccentricity of a binary increases, searches increasingly become blind to any potential population~\cite{Phukon:2024amh,Ramos-Buades:2020eju,Romero-Shaw:2025otx,Gadre:2024ndy}; this is compounded by the fact that signal-consistency tests which are required to handle non-Gaussian transient noise behavior~\cite{Allen:2004gu} will reject not just poorly fitted noise but also poorly fitted signals embedded in the data. This raises the question of whether the observation that most signals are consistent with low-eccentricity is truly indicative of the source population or if there could be a population of more highly eccentric binaries that have been overlooked.

In this work, we develop a new search method designed to identify eccentric NSBH binaries. The previous NSBH eccentric search~\cite{Dhurkunde:2023qoe} performed on data from the third observing run (O3) was already one of the most expensive analyses performed, requiring $\sim6$ million templates to span its search space; this is several times more expensive than the combined BNS/NSBH/BBH search performed for non-eccentric binaries by the LVK and others~\cite{LIGOScientific:2025zdk,Nitz:2021zwj}. This admittedly large cost though was only achieved by using a model that simplifies the gravitational-wave signal; while the model accurately handles the average dephasing caused by the burning off of the eccentric orbit~\cite{Moore:2016qxz}, it does make key simplifications which impact the ability to detect real sources. Namely, true eccentric orbits create harmonics in the gravitational wave signal and experience apsidal precession. Together, including these effects increases the sensitive search volume by $25\%\text{ to }30\%$. Waveform models now exist that include both these physical effects and in fact were used in an earlier BBH-targeted search~\cite{Wang:2025yac}. The use for NSBH searches has been limited by computational constraints; using the most accurate models increases the number of templates (and the cost) by just over an order of magnitude. In this paper, we develop a new method that combines multibanding~\cite{Adams:2015xza,Aubin:2020goo} with a hierarchical use of ratio-filtering~\cite{Nitz:2026fyr}, to speed up gravitational-wave searches by an order of magnitude; this makes this kind of large search possible for the first time. The resulting search covers nearly fifteen times as many templates as that previous analysis for comparable computational cost. We demonstrate the complete recipe required to execute a search of this size, including the generation of a template bank, the new optimized filtering approaches, the optimal handling of signal consistency tests, and a demonstration that this new method increases the sensitivity to eccentric binaries with $e_{20} \sim 0.28$ by $\sim 25\%\text{ to }30\%$. 

\begin{figure*}[!t]
    \centering
    \includegraphics[width=0.98\textwidth,height=0.95\textheight,keepaspectratio]{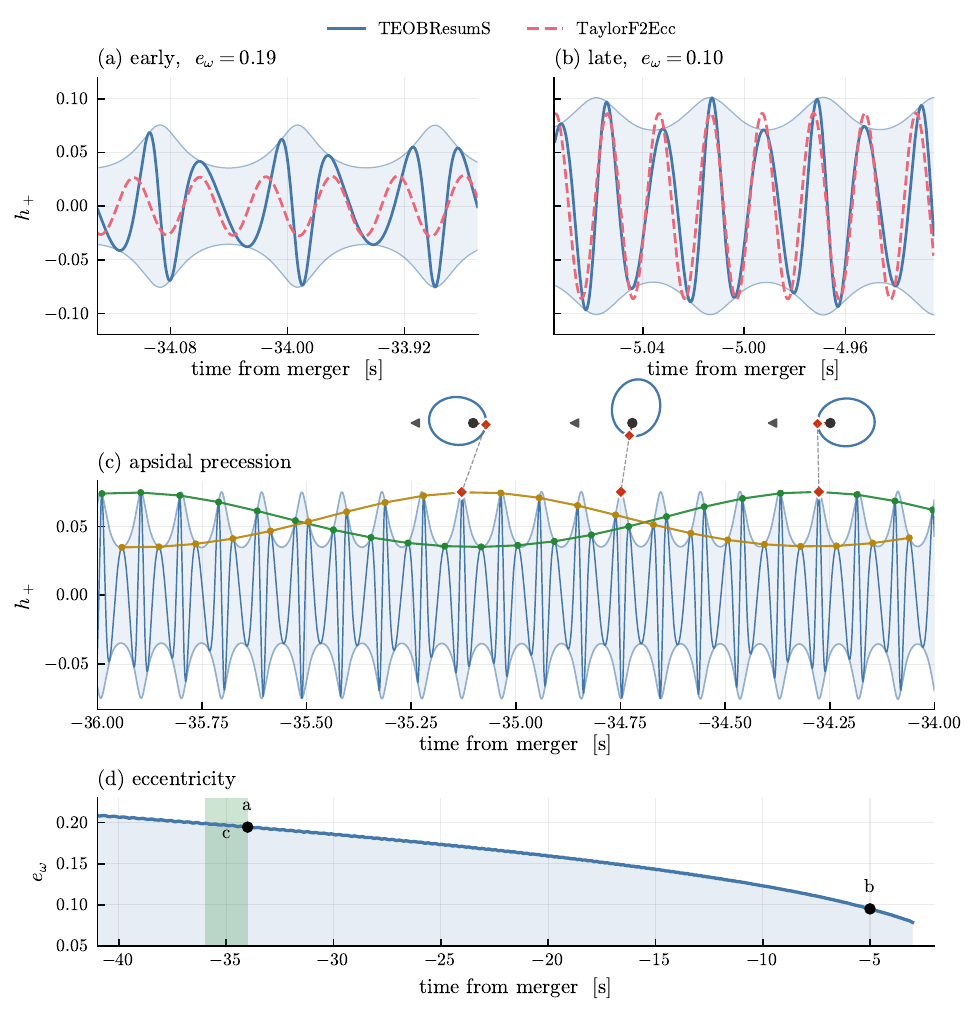}
    \caption{A time-domain comparison of the gravitational-wave strain $h_+$ produced by a moderately eccentric NSBH binary using the full-physics TEOBResumS model versus the dephasing-only TaylorF2Ecc approximant. Panels (a) and (b) highlight the early and late inspiral phases, illustrating the distinct inter-orbit harmonics present in the EOB model that are entirely absent in TaylorF2Ecc. Across these panels, the shaded blue envelopes denote the peak of the gravitational wave emission, providing a visual guide to compare the phasing from a particular viewing angle. Panel (c) isolates the amplitude modulations driven by apsidal precession: the yellow and green lines track the two vertices of the ellipse. The diagrammatic orbits above this panel illustrate the binary's orientation relative to an observer positioned to the left. Consequently, this panel effectively represents the strain as seen from an edge-on perspective, where only the plus polarization is observed. Panel (d) tracks the secular decay of the orbital eccentricity as the binary circularizes prior to merger. }
    \label{fig:harmonic}
\end{figure*}

\textit{Waveform Modelling and Computational Bottleneck.}---Previous searches for low-mass eccentric binaries used the frequency-domain approximant \texttt{TaylorF2Ecc}~\cite{Moore:2016qxz}, which captures the secular dephasing from orbital circularization~\cite{Peters:1964zz} but neglects apsidal precession and inter-orbit harmonics (Fig.~\ref{fig:harmonic}). Time-domain effective-one-body (EOB) models such as \texttt{TEOBResumS-DALI}~\cite{Nagar:2024dzj,Chiaramello:2020ehz,Nagar:2021gss} and \texttt{SEOBNRv5EHM}~\cite{Gamboa:2024hli,Gamboa:2024imd} include both effects and have been used in BBH searches~\cite{Wang:2025yac}, but their cost has so far prevented their use across the low-mass NSBH space. Including these effects has a negligible impact below $e_{20} \sim 0.1$, but increases the sensitive volume by $25\text{--}30\%$ for $0.2 \le e_{20} \le 0.28$.

Our templates use only the dominant $(2,2)$ mode, which preserves analytic maximization over orbital phase, inclination, and polarization. Neglecting higher-order modes reduces the detection rate by $\sim 5\%$ within our search region, mostly for edge-on sources (Appendix~\ref{app:waveforms}).

\textit{Rapid Multiband Ratio-Filtering Pipeline.}---Executing a wide-parameter matched-filter search with time-domain EOB waveforms is computationally prohibitive using standard direct filtering techniques. To make this eccentric NSBH search feasible, we build upon the \texttt{PyCBC} software library~\cite{Usman:2015kfa,Nitz:2018imz}, a framework utilized extensively in past gravitational-wave searches. We developed a rapid filtering architecture that integrates directly into the \texttt{PyCBC} search workflow system as a drop-in replacement for the standard \texttt{PyCBC Inspiral} search code. This new engine combines and extends two established acceleration methods: multibanding and ratio-filtering. Multibanding, originally introduced to gravitational-wave searches by the MBTA pipeline~\cite{Adams:2015xza,Aubin:2020goo,Allene:2025saz} and also similar to the multi-rate filtering employed by the GstLAL pipeline~\cite{Cannon:2011vi,Sachdev:2019vvd}, reduces computational cost by evaluating the low and high-frequency regimes of a signal at distinct sample rates. Ratio-filtering accelerates matched filtering by evaluating data against the slowly varying ratio between adjacent templates in a bank~\cite{Nitz:2026fyr}.

\begin{figure}[t]
    \centering
    \includegraphics[width=\linewidth]{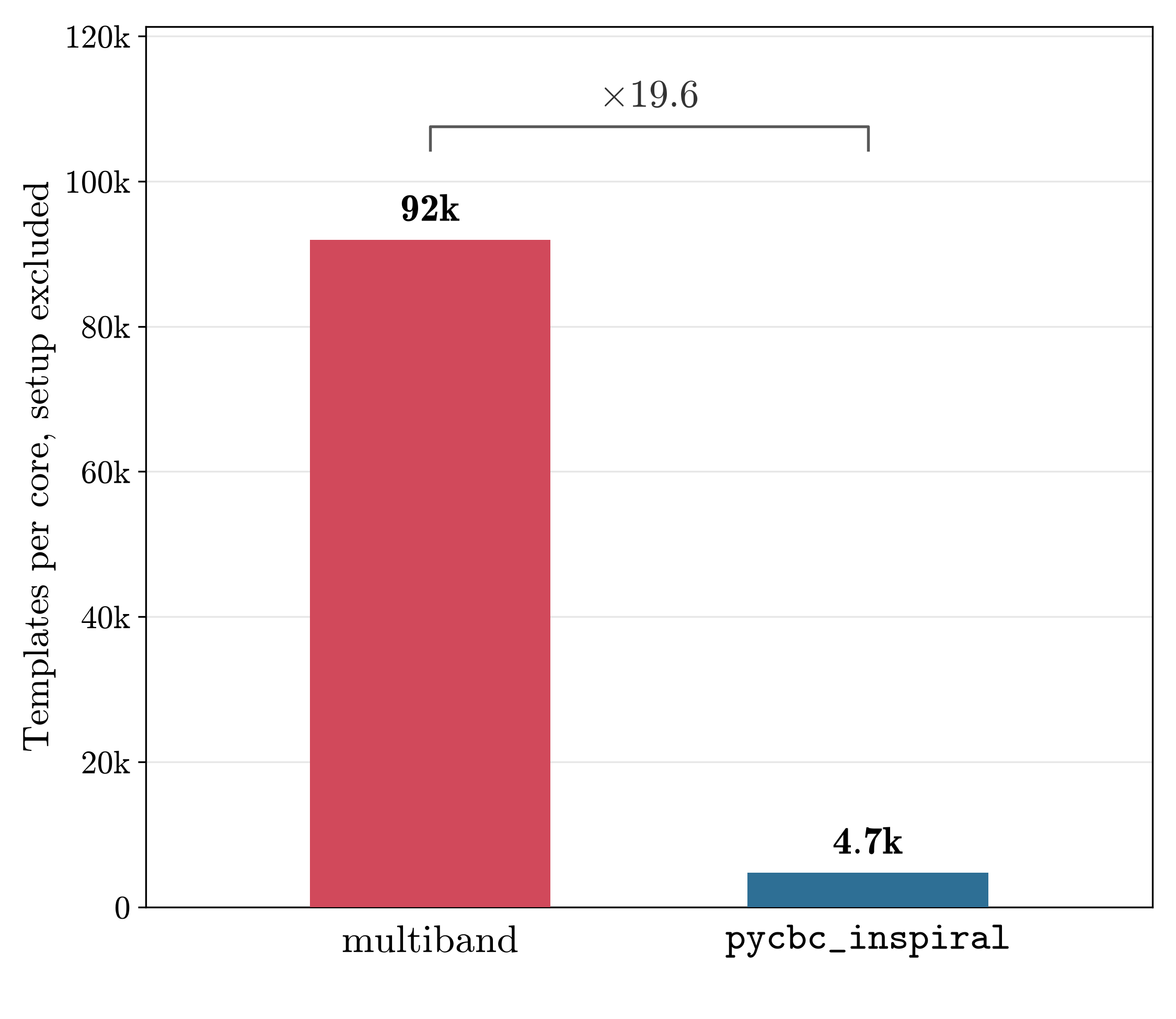}
    \caption{Computational throughput of the multiband pipeline compared with the standard
    \texttt{PyCBC Inspiral} code on the OrangeGrid High Throughput Computing (HTC)
    cluster, measured on one week of H1 and L1 data from O4a and normalized to
    single-core performance. The average cluster node uses $\sim 13$ year old Haswell CPUs. For a comparison on modern hardware, see Appendix~\ref{app:benchmarking}. Results are quoted in templates per core,
    excluding per-job setup time. The multiband bar is the mean of four analyses
    of the \texttt{TEOBResumS} bank (two injection campaigns in each of H1 and L1);
    the \texttt{PyCBC Inspiral} bar is the mean of eight analyses of the
    \texttt{TaylorF2} and \texttt{TaylorF2Ecc} banks over the same campaigns and
    detectors. All timings are taken from the injection-free analyses, so the
    analyses behind each bar are repeated measurements of the same quantity. The
    multiband pipeline achieves $\sim 20\times$ the throughput of
    \texttt{PyCBC Inspiral}.}
    \label{fig:cluster_throughput}
\end{figure}

Applying these methods to highly eccentric signals presents specific challenges. Standard ratio-filtering cannot be directly applied across the full bandwidth of eccentric waveforms. The presence of inter-orbit harmonics means there is no unique mapping between time and frequency, preventing the template ratios from cleanly dividing into a compact form. However, because these harmonics decay as the binary circularizes, the ratio-filtering approach remains highly effective in the high-frequency regime. To capitalize on this, we implement a hierarchical triggering scheme. The data is first filtered against the low-frequency portion of the templates. When a trigger exceeds a preliminary signal-to-noise ratio threshold in this band, the peak is temporally interpolated to a higher sample rate, and the high-frequency ratio-filtered contribution is dynamically calculated and added. By integrating this hierarchical multibanding approach with targeted ratio-filtering and a rapid waveform reconstruction algorithm (detailed in Appendices~\ref{app:compression} and~\ref{app:multiband}), the resulting pipeline achieves a $\sim 20\times$ reduction in computational cost while preserving the full physics of the EOB models.

To assess performance under realistic operating conditions, we deployed the full search workflow on the OrangeGrid High Throughput Computing cluster, analyzing one week of H1 and L1 data from O4a. We performed injection campaigns with two sets of simulated signals generated with \texttt{TEOBResumS}, both including higher-order modes: one restricted to aligned spins and one with precessing spins. Each injection set was analyzed with three template banks: a \texttt{TEOBResumS} bank restricted to the dominant $(2,2)$ mode containing $\sim 5\times10^{4}$ templates, a \texttt{TaylorF2Ecc} bank of 734 templates, and a \texttt{TaylorF2} bank of 487 templates. 

Figure~\ref{fig:cluster_throughput} shows the throughput measured in these production analyses. The multiband pipeline processes $\sim 9.2\times10^{4}$ templates per core, compared with $\sim 4.7\times10^{3}$ for \texttt{PyCBC Inspiral}, an improvement of $\sim 20\times$. This exceeds the $8.9\times$ gain measured in controlled single-node benchmarks (see Appendix~\ref{app:benchmarking}). The larger factor likely reflects the heterogeneous hardware and variable load of a shared production cluster. As contention increases, the reduced memory footprint of the multiband pipeline makes more efficient use of the hardware memory hierarchy. The production measurement confirms that this speedup is retained, and in practice enhanced, under realistic operating conditions.

Signal consistency tests, such as the $\chi^2$ discriminator, are adapted to this architecture by evaluating the multi-bin statistic across the low-frequency band while treating the high-frequency ratio band as a single additive bin. Paired with a descending-SNR evaluation heuristic (detailed in Appendix~\ref{app:consistency}), the test successfully separates true astrophysical signals from non-Gaussian background noise across real detector data up to SNR $\sim 100$.

\textit{Observational Search and Candidate Analysis.}---We deployed the complete hierarchical multiband search workflow across the first two observing epochs of the fourth observing run (O4a and O4b) of Advanced LIGO, analyzing data from the Hanford (H1) and Livingston (L1) observatories spanning both the O4a and O4b observing epochs. The search covers both coincident and non-coincident (single-detector) observing time, analyzing $238\ \mathrm{days}$ ($0.65\ \mathrm{yr}$) of coincident H1L1 observing time as well as $163\ \mathrm{days}$ ($0.45\ \mathrm{yr}$) of non-coincident single-detector active time ($56\ \mathrm{days}$ in H1 and $107\ \mathrm{days}$ in L1), yielding a total search baseline of $401\ \mathrm{days}$ ($1.10\ \mathrm{yr}$). The targeted parameter space spans component masses $m_1 \in [3, 15]\,M_\odot$ and $m_2 \in [1.2, 2.2]\,M_\odot$, aligned primary spin $|\chi_{1z}| \le 0.5$, secondary spin $|\chi_{2z}| \le 0.1$, and initial orbital eccentricity at $20\ \mathrm{Hz}$ extending up to $e_{20} = 0.28$. To manage the computational scale, the space is partitioned into 50 parallel subsearches containing a combined 88.6~million EOB templates. Because 50 subsearches are evaluated independently over the data, we apply a global trials factor of 50 to all false alarm rates ($\mathrm{FAR}_{\mathrm{trials}} \approx 50 \times \mathrm{FAR}_{\mathrm{subsearch}}$) when assigning statistical significance.

\begin{table}[t]
\caption{Candidate events and loudest foreground triggers in the O4a and O4b search across both coincident and non-coincident observing time. Stat represents the standard multi-detector PyCBC ranking statistic~\cite{Davies:2020tsx}. Single-detector IFAR values are estimated via fitted background extrapolations \cite{Davies:2020tsx}, capped at $>100$~years. The IFAR of GW230518\_125908 is bounded by the background livetime. Both GW230518\_125908 and GW230529\_181500 were recovered with parameters consistent with zero eccentricity ($e_{20} \to 0$). Masses for the two known events are source-frame GWTC-4.0 medians and 90\% credible intervals~\cite{LIGOScientific:2025slb}; all other columns, and the masses for GPS 1377876363, are from the best-matching search template (detector frame). Triggers failing data quality cuts and localized glitch vetoes are excluded. The loudest remaining noise trigger in coincident time has $\mathrm{IFAR} \approx 0.12$~yr, consistent with noise; the trials factor of 50 is conservative, as neighboring subsearches overlap and their triggers are correlated.}
\label{tab:candidates}
\begin{ruledtabular}
\setlength{\tabcolsep}{3pt}
\begin{tabular}{lccccc}
Trigger & $m_1$ & $m_2$ & $e_{20}$ & Stat & IFAR [yr] \\
 & \multicolumn{2}{c}{$[M_\odot]$} & & & \\
\hline
GW230518\_125908 & $8.2^{+0.8}_{-0.9}$ & $1.5^{+0.1}_{-0.1}$ & 0.02 & 43.9 & $>100$ \\
GW230529\_181500\footnote{Single-detector trigger in L1 (H1 offline).} & $3.7^{+0.8}_{-1.2}$ & $1.4^{+0.6}_{-0.2}$ & 0.00 & 27.5 & $>100$ \\
1377876363 & 8.3 & 1.4 & 0.20 & 16.0 & $0.12$ \\
\end{tabular}
\end{ruledtabular}
\end{table}

Table~\ref{tab:candidates} summarizes the candidate gravitational-wave events and loudest foreground triggers identified by the search. As in standard search methodology, triggers failing automated data-quality vetoes and localized non-Gaussian burst vetoes are discarded. The pipeline cleanly recovers the two known low-mass coalescences observed during O4: GW230518\_125908 ($m_1 \approx 8.2\,M_\odot, m_2 \approx 1.45\,M_\odot$~\cite{LIGOScientific:2025slb}) at overwhelming significance ($\mathrm{Stat} = 43.9$, louder than all time-slide background, so its IFAR is bounded by the background livetime), and GW230529\_181500 ($m_1 \approx 3.7\,M_\odot, m_2 \approx 1.4\,M_\odot$, $\mathrm{Stat} = 27.5$) during non-coincident single-detector L1 time. Both events are recovered with templates possessing near-zero initial eccentricity ($e_{20} \le 0.02$), consistent with quasicircular inspirals. The primary of GW230529\_181500 lies within the lower mass gap and may be a low-mass black hole or a massive neutron star, but neither event exhibits evidence for non-zero eccentricity. After discarding vetoed triggers, the loudest remaining foreground trigger across the search occurs at GPS 1377876363 with a ranking statistic of $\mathrm{Stat} = 16.0$ ($\mathrm{IFAR} = 0.12\ \mathrm{yr}$), consistent with noise given the conservative trials factor. No significant eccentric candidates were identified in either coincident or single-detector data. We conclude that no new gravitational-wave events with non-zero orbital eccentricity ($e_{20} \ge 0.05$) were detected.

\begin{figure}[t]
    \centering
    \includegraphics[width=\linewidth]{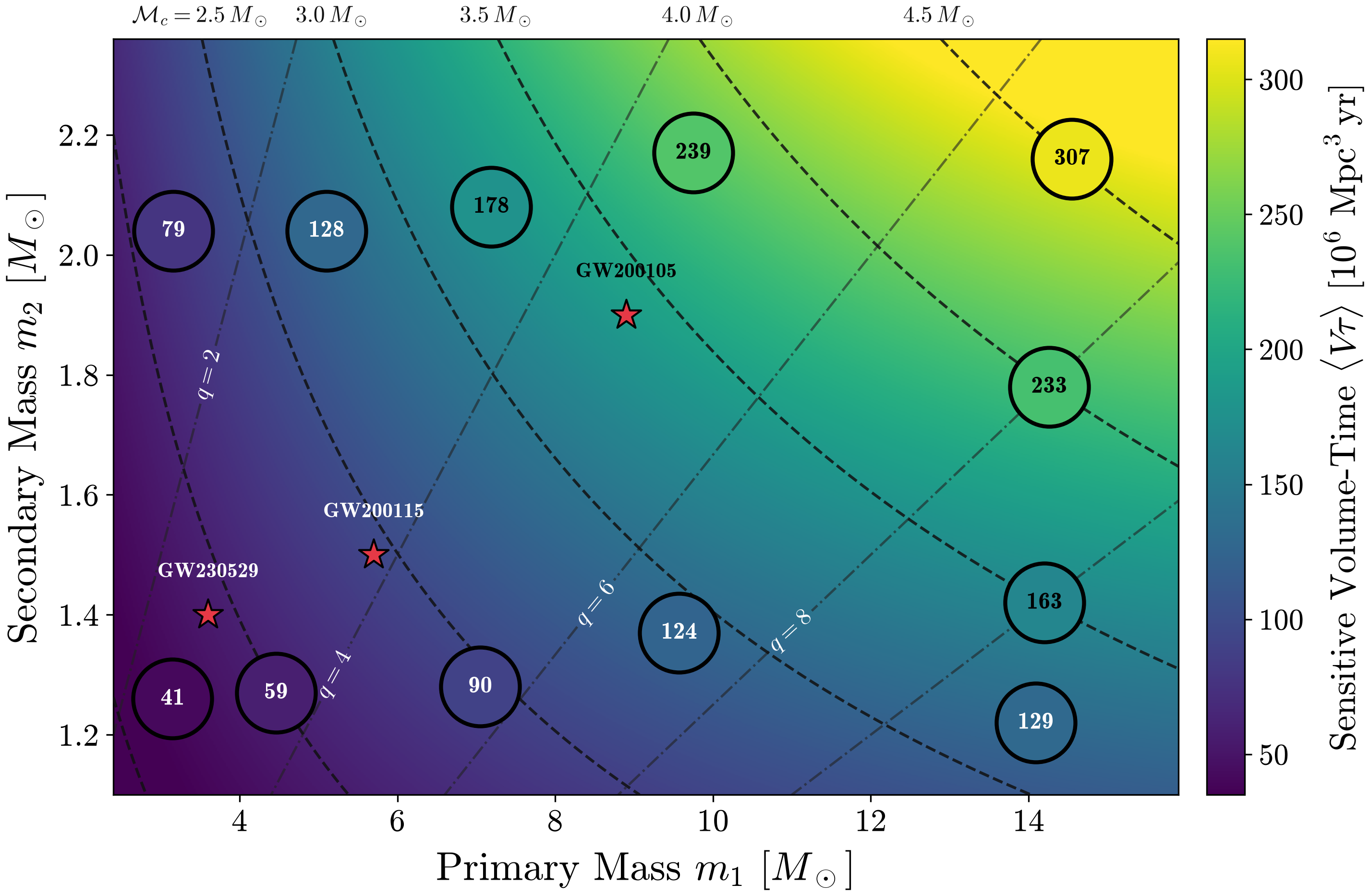}
    \caption{Sensitive volume-time $\langle V\tau \rangle\ [10^6\ \mathrm{Mpc}^3\,\mathrm{yr}]$ across the target NSBH parameter space in detector-frame component masses $(m_1, m_2)$ for $e_{20} \le 0.28$. Numbered circles show the evaluated injection ensembles, while the background color map shows the fitted physical scaling across the plane. Dashed lines trace contours of constant chirp mass $\mathcal{M}_c$, and dash-dotted lines indicate constant mass ratios $q = m_1 / m_2$. Red star markers locate three known LVK events in this mass vicinity: GW230529\_181500, GW200115\_042309, and GW200105\_162426.}
    \label{fig:m1m2_sensitivity}
\end{figure}

\textit{Search Sensitivity and Astrophysical Constraints.}---We quantify the astrophysical reach of the search across the targeted parameter space using an extensive campaign of simulated signals. Figure~\ref{fig:m1m2_sensitivity} maps the sensitive volume-time $\langle V\tau \rangle$ across the component mass plane. Across the targeted space ($e_{20} \le 0.28$), $\langle V\tau \rangle$ ranges from $\sim 40 \times 10^6\ \mathrm{Mpc}^3\,\mathrm{yr}$ for low-mass systems ($m_1 \sim 3\,M_\odot, m_2 \sim 1.3\,M_\odot$, corresponding to an effective survey radius $R_{\mathrm{eff}} \sim 210\ \mathrm{Mpc}$) up to $\sim 310 \times 10^6\ \mathrm{Mpc}^3\,\mathrm{yr}$ for the most massive black hole-neutron star systems ($m_1 \sim 15\,M_\odot, m_2 \sim 2.2\,M_\odot$, $R_{\mathrm{eff}} \sim 410\ \mathrm{Mpc}$). Within the template bank boundaries ($e_{20} \le 0.28$, aligned spins), we find minimal variation in search sensitivity as a function of eccentricity, allowing the sensitivity map to be compactly parameterized by $m_1$ and $m_2$.

\begin{figure}[t]
    \centering
    \includegraphics[width=\linewidth]{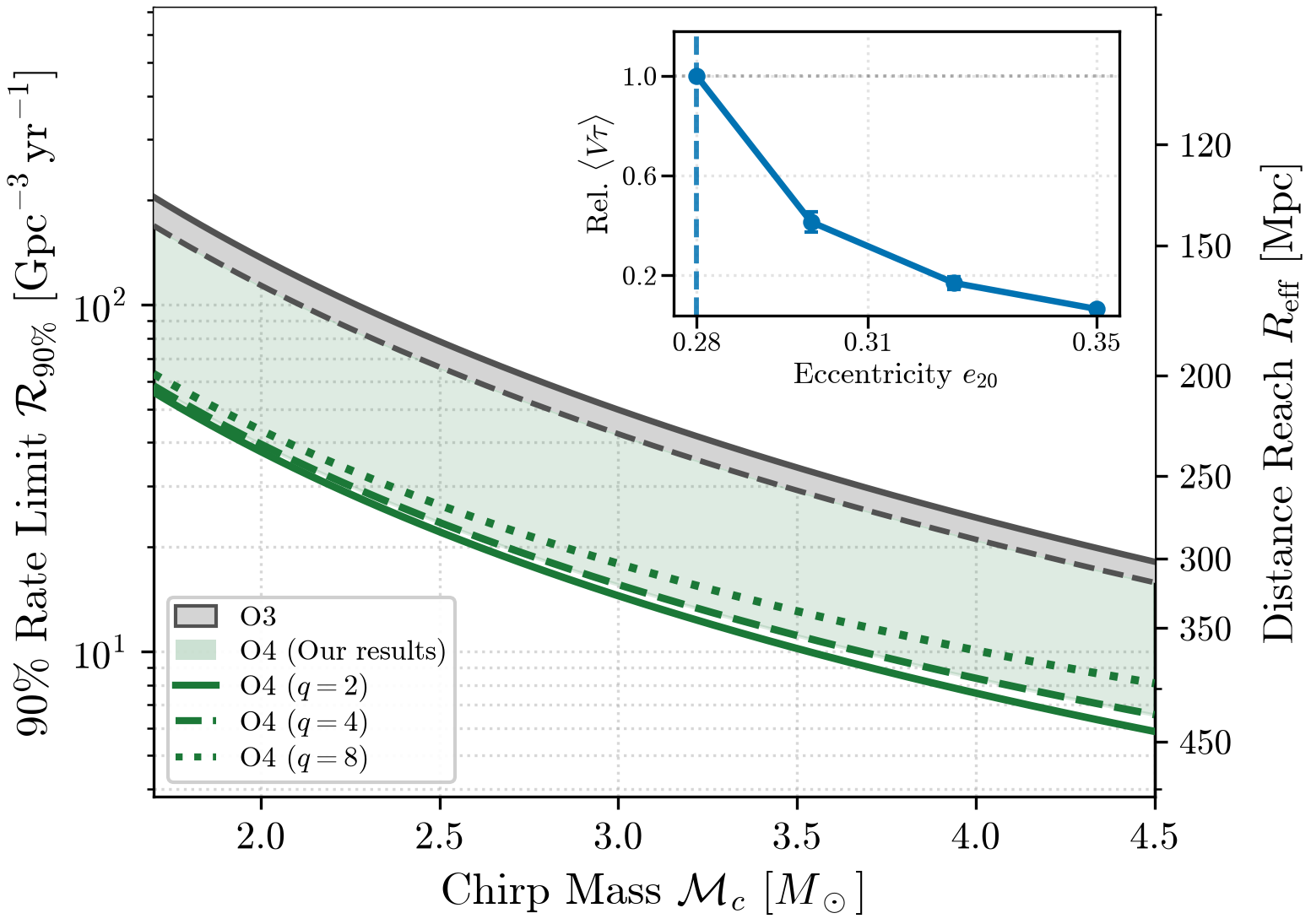}
    \caption{The 90\% credible merger rate density upper limits $\mathcal{R}_{90\%}$ (left axis) and effective sensitive distance reach $R_{\mathrm{eff}}$ (right axis) from the O4a and O4b search as a function of chirp mass $\mathcal{M}_c$, compared against the prior O3 analysis of Dhurkunde et al.~\cite{Dhurkunde:2023qoe}. Green curves present our O4 limits for mass ratios $q \in \{2, 4, 8\}$ ($m_1/m_2$), with the shaded green region denoting the newly ruled-out parameter space relative to O3. Gray curves bound the O3 search limits spanning $e_{20} \to 0$ (dashed) up to $e_{20} \sim 0.28$ (solid, rescaled by the $25\text{--}30\%$ volume gain of EOB templates over \texttt{TaylorF2Ecc}), with the shaded gray band indicating the O3 range. Across the parameter space, our search achieves up to a $\sim 4\times$ improvement over previous observational constraints. The upper-right inset shows the relative sensitive volume-time $\langle V\tau \rangle$ drop-off beyond the template bank boundary as a function of orbital eccentricity $e_{20}$ (normalized to the sensitivity at the template bank boundary $e_{20}=0.28$, marked by the vertical dashed line).}
    \label{fig:astrophysical_limits}
\end{figure}

Because no eccentric NSBH mergers were identified in the data, we place $90\%$ credible upper limits on the local merger rate density of eccentric NSBH systems using the loudest event statistic~\cite{Biswas:2007ni}:
\begin{equation}
\mathcal{R}_{90\%} = \frac{2.303}{\langle V\tau \rangle},
\end{equation}
where $\langle V\tau \rangle$ is the sensitive volume-time accumulated across O4 evaluated at the ranking statistic of the loudest foreground event ($\mathrm{Stat} = 16.0$). Figure~\ref{fig:astrophysical_limits} displays these upper limits and the corresponding effective distance reach $R_{\mathrm{eff}}$, benchmarked against the O3 search of Dhurkunde et al.~\cite{Dhurkunde:2023qoe}. The resulting constraints span $\mathcal{R}_{90\%} \sim 7.5\text{--}57\ \mathrm{Gpc}^{-3}\ \mathrm{yr}^{-1}$ ($R_{\mathrm{eff}} \sim 210\text{--}410\ \mathrm{Mpc}$), representing an improvement of up to a factor of $\sim 4$. This gain is driven jointly by the improved detector sensitivity of O4 and the physical volume recovery of the EOB waveforms, which increase the sensitive volume by $25\%\text{ to }30\%$ relative to simplified models like \texttt{TaylorF2Ecc} at $e_{20} \sim 0.28$ (see Appendix~\ref{app:relative_vt}). In contrast, standard circular template searches miss roughly $80\%$ of the sensitive volume recovered by our EOB bank at these eccentricities. As shown in the inset of Fig.~\ref{fig:astrophysical_limits}, the template bank retains sensitivity even outside its nominal construction region, recovering $\sim 40\%$ of the in-band sensitive volume at $e_{20} = 0.30$ ($\sim 7\%$ by $e_{20} = 0.35$). Using a separate set of simulated signals with precessing spins (component spin magnitudes up to $0.3$, $\chi_p \lesssim 0.3$), we find that, assuming an isotropic distribution of spin orientations, the loss in sensitivity due to spin precession is small within this spin range.

At the specific parameters of GW200105\_162426 ($m_1 \approx 8.9\,M_\odot, m_2 \approx 1.9\,M_\odot, \mathcal{M}_c \approx 3.41\,M_\odot$), under the standard null hypothesis where GW200105\_162426 is treated as non-eccentric and no eccentric NSBH mergers have occurred, our non-detection sets a $90\%$ credible upper limit of $\mathcal{R}_{90\%} \le 12.1\ \mathrm{Gpc}^{-3}\ \mathrm{yr}^{-1}$ from O4. Conversely, motivated by recent suggestions of residual orbital eccentricity in GW200105\_162426~\cite{Morras:2025xfu,Planas:2025plq,Kacanja:2025kpr}, we can consider the hypothesis that GW200105\_162426 was the first detection from an eccentric NSBH subpopulation. Treating GW200105\_162426 as a single observed event ($k=1$), and combining our O4 sensitivity with that of the O3 eccentric search of Dhurkunde et al.~\cite{Dhurkunde:2023qoe}, the absence of further detections in O4 updates the inferred local merger rate density to a median of $\mathcal{R} = 6.6^{+6.3}_{-3.8}\ \mathrm{Gpc}^{-3}\ \mathrm{yr}^{-1}$ ($68\%$ credible interval), with a $90\%$ credible upper bound of $\mathcal{R}_{90\%} \le 15.3\ \mathrm{Gpc}^{-3}\ \mathrm{yr}^{-1}$. These limits are comparable to the overall NSBH merger rate density inferred from quasicircular analyses ($\sim 6.7\text{--}32.8\ \mathrm{Gpc}^{-3}\ \mathrm{yr}^{-1}$~\cite{LIGOScientific:2026ctl}).

\textit{Conclusions.}---The observation of eccentric compact binaries would provide direct evidence for the dynamical formation of neutron star-black hole systems. However, the computational cost of evaluating time-domain effective-one-body waveforms across a wide parameter space has historically forced searches to rely on simplified models. These models neglect inter-orbit harmonics and apsidal precession; including these physical effects increases the sensitive volume for highly eccentric sources ($e_{20} \sim 0.28$) by $\sim 25\%$ to $30\%$.

In this work, we presented a rapid matched-filter methodology that makes the direct use of these waveforms computationally feasible. By combining sparse waveform compression, discrete multibanding, and hierarchical high-frequency ratio-filtering, the pipeline achieves a $\sim 20\times$ speedup over standard FFT-based matched filtering on high-throughput clusters. We deployed this architecture across the O4a and O4b observing epochs of Advanced LIGO, searching an 88.6-million template bank of EOB waveforms over 401~days of coincident and non-coincident data (238~days of coincident H1L1 time and 163~days of single-detector time). We recovered known low-mass mergers GW230518\_125908 and GW230529\_181500 with parameters consistent with zero eccentricity. Finding no new gravitational-wave events with non-zero eccentricity, we established the deepest rate density upper limits to date on eccentric NSBH mergers ($\mathcal{R}_{90\%} \sim 7.5\text{--}57\ \mathrm{Gpc}^{-3}\ \mathrm{yr}^{-1}$), tightening observational constraints by up to a factor of $\sim 4$.

This architecture is integrated directly into the open-source \texttt{PyCBC} framework, demonstrating that gravitational-wave searches can fully leverage the most accurate available waveform models without being bottlenecked by computational constraints and paving the way for future wide-parameter analyses in the next generation of detectors.

\section*{Data Availability}
The data and scripts needed to reproduce the figures in this work are available at \url{https://github.com/sugwg/eccentric-nsbh-search}. The hierarchical multiband ratio-filtering implementation described in this work is available in the \texttt{multiband} branch of \url{https://github.com/ahnitz/pycbc}, and is intended for inclusion in a future release of the mainline \texttt{PyCBC} distribution.

\begin{acknowledgments}
We acknowledge the support of Syracuse University for providing the computational resources through the OrangeGrid High Throughput Computing (HTC) cluster supported by the NSF award ACI-1341006. This research was done using services provided by the OSG Consortium~\cite{pordes2007open,sfiligoi2009pilot,osg2006ospool,osg2015openscience} which is supported by the National Science Foundation awards \#2030508 and \#2323298.

This research has made use of data or software obtained from the Gravitational Wave Open Science Center (gwosc.org), a service of the LIGO Scientific Collaboration, the Virgo Collaboration, and KAGRA. This material is based upon work supported by NSF's LIGO Laboratory which is a major facility fully funded by the National Science Foundation, as well as the Science and Technology Facilities Council (STFC) of the United Kingdom, the Max-Planck-Society (MPS), and the State of Niedersachsen/Germany for support of the construction of Advanced LIGO and construction and operation of the GEO600 detector. Additional support for Advanced LIGO was provided by the Australian Research Council. Virgo is funded, through the European Gravitational Observatory (EGO), by the French Centre National de Recherche Scientifique (CNRS), the Italian Istituto Nazionale di Fisica Nucleare (INFN) and the Dutch Nikhef, with contributions by institutions from Belgium, Germany, Greece, Hungary, Ireland, Japan, Monaco, Poland, Portugal, Spain. KAGRA is supported by Ministry of Education, Culture, Sports, Science and Technology (MEXT), Japan Society for the Promotion of Science (JSPS) in Japan; National Research Foundation (NRF) and Ministry of Science and ICT (MSIT) in Korea; Academia Sinica (AS) and National Science and Technology Council (NSTC) in Taiwan.
\end{acknowledgments}

\bibliography{main}

\clearpage
\appendix

\section{Eccentric Binary Waveform Models and Physical Approximations}
\label{app:waveforms}

The sensitivity of a matched-filter search depends directly on the phase and amplitude accuracy of the underlying waveform templates. Previous searches for low-mass eccentric binaries primarily utilized the frequency-domain approximant \texttt{TaylorF2Ecc}~\cite{Moore:2016qxz}. This model captures the effective secular dephasing induced by the emission of gravitational radiation and the resulting circularization of the orbit~\cite{Peters:1964zz,Moore:2016qxz}. It simplifies the signal morphology by neglecting the amplitude modulations driven by apsidal precession and omitting the inter-orbit harmonics naturally generated by eccentric motion. These simplifications are sufficiently accurate for low-eccentricity sources ($e_{20} < 0.1$), but they introduce a moderate loss in sensitivity as initial eccentricity increases~\cite{Ramos-Buades:2020eju,Gadre:2024ndy}.

\begin{figure*}[t]
    \centering
    \includegraphics[width=\linewidth]{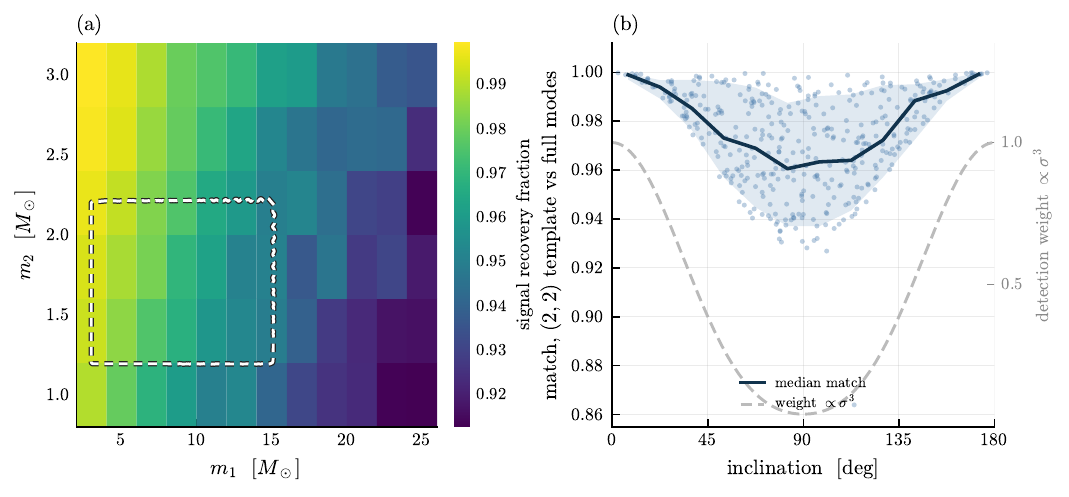}
    \caption{The fractional loss in signal recovery when filtering a complete multi-modal eccentric waveform with a template restricted strictly to the dominant (2,2) mode. Panel (a) displays the signal recovery fraction as a function of the detector-frame redshifted component masses. The dotted line delineates the boundary of the specific search space targeted in this work. Even within this defined search region, neglecting higher-order modes results in an approximate 5\% loss in the overall detection rate. Panel (b) illustrates the match and detection weight as a function of orbital inclination, demonstrating that this sensitivity loss is predominantly driven by edge-on source configurations (approaching 90 degrees) where the contribution of higher-order modes is most pronounced.}
    \label{fig:highermode}
\end{figure*}

To capture the full dynamics of highly eccentric mergers, modern time-domain effective-one-body (EOB) models such as \texttt{TEOBResumS-DALI}~\cite{Nagar:2024dzj,Chiaramello:2020ehz,Nagar:2021gss} and \texttt{SEOBNRv5EHM}~\cite{Gamboa:2024hli,Gamboa:2024imd} are required. These models have been deployed in prior searches targeting binary black holes~\cite{Wang:2025yac}. Their use in BNS and NSBH searches has historically been prohibited by the sheer size of the low-mass parameter space and the associated computational scaling. In addition to EOB models, other waveform families like \texttt{Esigma}~\cite{Paul:2024ujx} and the inspiral-only \texttt{PyEFPE}~\cite{Morras:2025nlp} model provide alternative frameworks for capturing eccentric dynamics. As illustrated in Fig.~\ref{fig:harmonic}, the omission of harmonics and apsidal precession in simpler models leads to a measurable dephasing and amplitude mismatch during the early inspiral. The gain in sensitive volume from including them is negligible for eccentricities below $0.1$ at a reference frequency of $20\mathrm{\ Hz}$, but grows to roughly $25\text{--}30\%$ for $0.2 \le e_{20} \le 0.28$.

Beyond eccentricity-driven harmonics, these highly accurate models natively include higher-order modes (HOMs). We restrict our template bank generation to the dominant $(2,2)$ mode. This restriction preserves the algebraic decomposition that allows for analytical maximization over the orbital phase, inclination, and polarization. We numerically confirmed that this decoupling holds robustly in practice for the EOB models utilized in this search. There are existing methods designed to make HOMs computationally feasible in matched-filter searches~\cite{Harry:2017weg,Schmidt:2024jbp}, and their inclusion remains a target for future development. As shown in Fig.~\ref{fig:highermode}, excluding HOMs causes a reduced signal recovery fraction for edge-on source orientations. This reduces sensitivity to systems that are particularly scientifically valuable for measuring effects like spin precession. Assuming an isotropic source population, however, the HOM contribution remains subdominant on average. A $(2,2)$-mode restricted template recovers the bulk of the eccentric NSBH population while avoiding the additional structural complexity of a multi-modal search.

\section{Stochastic Template Bank Construction and Parameter Space Coverage}
\label{app:bank}

Conducting a wide-parameter matched-filter search requires a template bank that densely covers the targeted physical parameter space. We employ a stochastic template placement algorithm~\cite{Harry:2009ea,Ajith:2012mn,Kacanja:2024pjh} because the underlying code is general and readily extends to higher dimensions. While geometric placement methods exist for eccentric searches~\cite{Phukon:2024amh}, they require an analytical metric of the parameter space. Currently available metrics do not account for the eccentric harmonics and apsidal precession present in full-physics EOB models. The stochastic placement naturally bypasses this limitation because it does not require knowing the metric ahead of time. In addition to component masses $m_1, m_2$ and aligned spin components ($|\chi_{1z}| \le 0.5$ for the primary and $|\chi_{2z}| \le 0.1$ for the secondary), there are two extra parameters in this search: orbital eccentricity at a reference frequency of 20~Hz and an anomaly parameter. Because we restrict the templates to the $(2,2)$ mode, orbital inclination and orbital phase remain degenerate and do not expand the dimensionality of the template bank. We set the target minimal match for the bank to $0.95$.

\begin{figure*}[t]
    \centering
    \includegraphics[width=\textwidth]{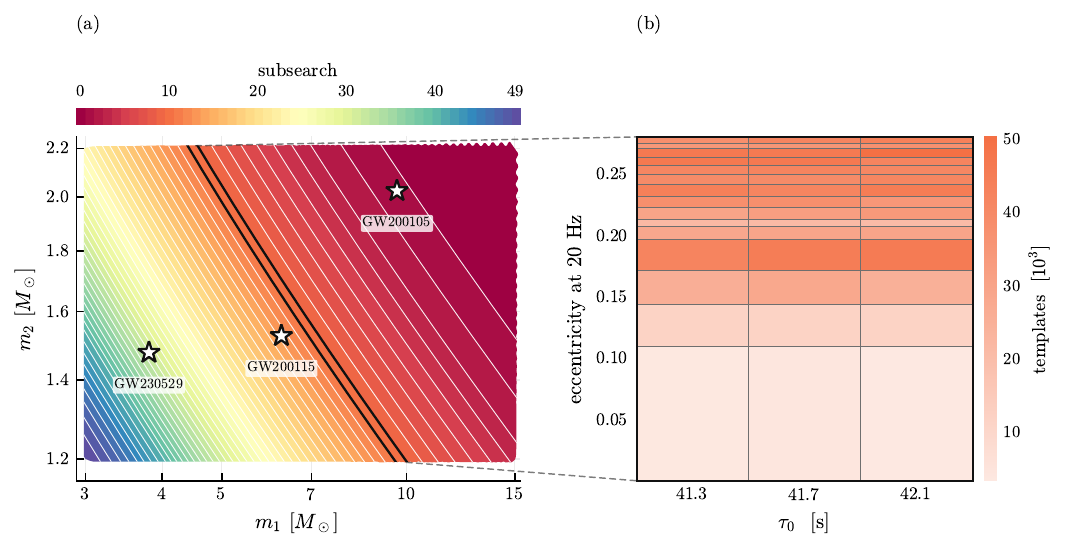}
\caption{The distribution of templates across the target parameter space. Panel (a) displays the search region in component masses $m_1$ and $m_2$, divided into distinct subsearches. The approximate locations of relevant gravitational-wave events (GW230529\_181500, GW200115\_042309, and GW200105\_162426) are overlaid. Panel (b) illustrates that each subsearch is further subdivided into discrete two-dimensional sub-banks bounded by chirp time $\tau_0$ and initial eccentricity at 20~Hz. The color scale denotes the number of templates (in thousands) required to cover each localized bin.}
    \label{fig:bank_space}
\end{figure*}

\begin{figure*}[t]
    \centering
    \includegraphics[width=\textwidth]{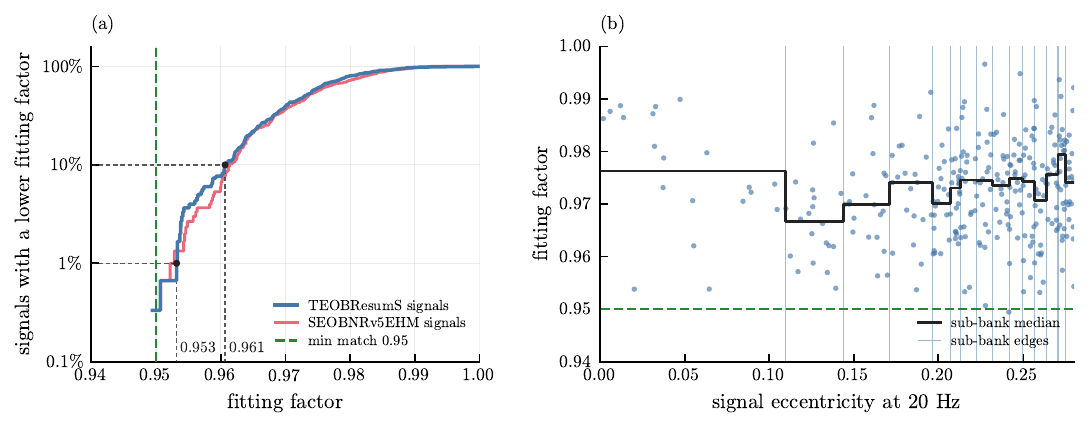}
\caption{The fitting factor recovery of the generated template bank. Panel (a) shows the cumulative distribution of the fitting factor for a simulated population of signals, evaluated independently against the \texttt{TEOBResumS} and \texttt{SEOBNRv5EHM} waveform families. The vertical dashed line indicates the target minimal match of 0.95. Panel (b) plots the fitting factor as a function of the injected signal's eccentricity at 20~Hz. Both individual signal recoveries and the sub-bank median are shown to demonstrate that the placement scheme satisfies the minimal match criterion across the targeted eccentricity range.}
    \label{fig:ff_subsearch}
\end{figure*}

\begin{figure}[t]
    \centering
    \includegraphics[width=\linewidth]{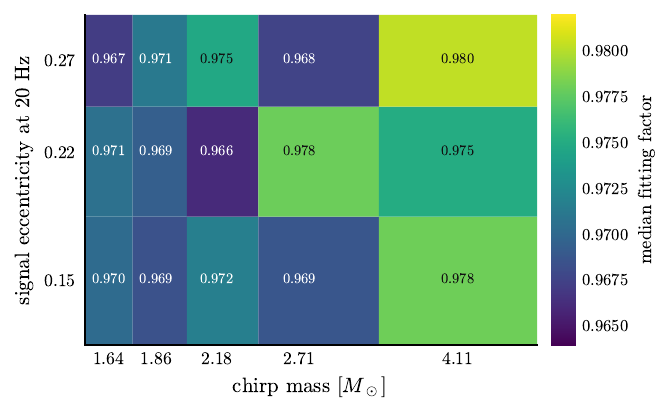}
\caption{A heatmap illustrating the median fitting factor of the template bank evaluated across the parameter space, binned by chirp mass and signal eccentricity at 20~Hz. The stochastic placement maintains a median fitting factor above 0.965 across all subdivided regions. This indicates that the discreteness of the template bank is expected to cause less than a $10\%$ loss in the overall signal recovery rate.}
    \label{fig:ff_bank}
\end{figure}

We generate the template bank in parallel by subdividing the parameter space into localized two-dimensional bins over chirp time $\tau_0$ and eccentricity at 20~Hz. The parallel construction creates over coverage at boundaries which we estimate to be $15-20\%$. We verified this by constructing a bank that covers the union of two adjacent spaces, testing this in a few representative locations throughout the bank. There is also intrinsic overcoverage resulting from the stochastic placement method itself, an expected feature detailed in recent comparisons of random template banks and geometric lattices~\cite{Kacanja:2024pjh}. The computational cost of bank construction remains negligible compared to the continuous cost of filtering data, however, increased template count does scale the total search cost.

Figure~\ref{fig:bank_space} illustrates the boundaries of our subdivided search space in component mass alongside the distribution of templates across the $\tau_0$ and eccentricity grid. Figure~\ref{fig:ff_subsearch} demonstrates that the resulting bank maintains a high minimal match when evaluated against both the \texttt{TEOBResumS-DALI} and \texttt{SEOBNRv5EHM} waveform families. This confirms the placement scheme is robust against inherent waveform systematics between these two models in this particular region of the parameter space, though we note that waveform systematics have been observed in other areas, such as the BBH parameter space~\cite{Gamboa:2024hli}. Figure~\ref{fig:ff_bank} confirms the global fitting factor recovery across chirp mass and initial eccentricity, showing the stochastic placement ensures targeted signals are recovered efficiently.

\section{Waveform Compression and Rapid Nodal Reconstruction}
\label{app:compression}

Generating time-domain EOB waveforms for long-duration NSBH signals is computationally expensive, typically requiring seconds to tens of seconds per waveform. While on-the-fly generation is acceptable for localized injection studies, a wide-parameter matched-filter search requires evaluating data against tens of millions of templates. Precomputing and directly storing the full-resolution time-domain templates is unfeasible due to excessive storage requirements. To resolve this, we employ a reduced-representation approach where the waveform is compressed into a sparse array of amplitude and phase nodes, stored on disk, and dynamically reconstructed in memory during the search. To achieve this, the time-domain EOB signal is generated, a short taper is applied to the start, and the signal is Fourier transformed. We then apply the compression scheme, which was originally designed for analytic frequency-domain waveforms, directly to this numerical frequency-domain output.

\begin{figure*}[t]
    \centering
    \includegraphics[width=\textwidth]{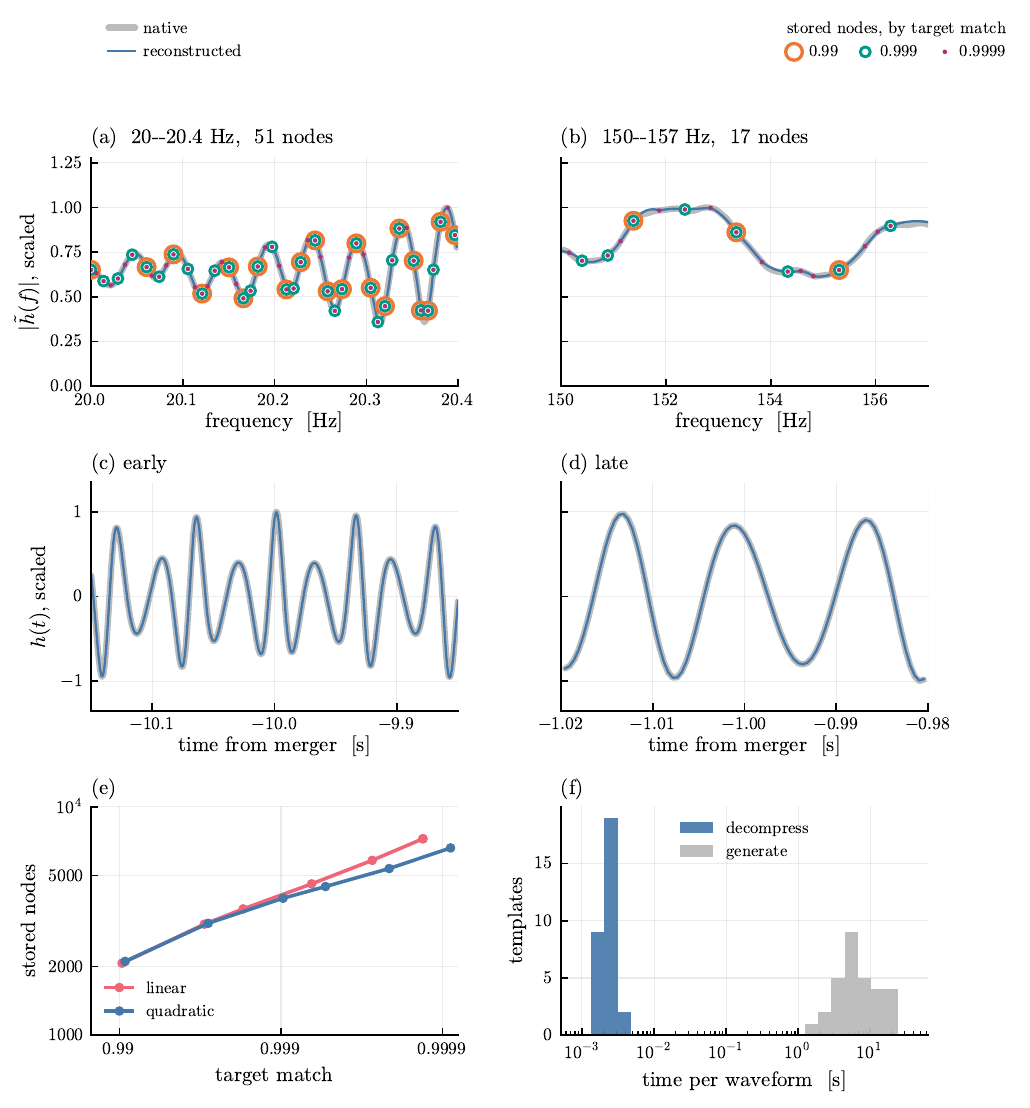}
    \caption{The temporal compression and reconstruction scheme for eccentric time-domain waveforms. Panels (a) and (b) show the native frequency-domain amplitude alongside the sparsely populated reconstruction nodes for the early ($20\text{ to }20.4\text{ Hz}$) and late ($150\text{ to }157\text{ Hz}$) inspiral. Panels (c) and (d) overlay the reconstructed time-domain strain on the native signal. Panel (e) compares the number of stored nodes required to achieve a specific target match using linear versus quadratic interpolation, demonstrating that the quadratic scheme requires fewer nodes at high accuracies. Panel (f) displays a histogram of the processing time per waveform. The dynamic decompression from the stored nodes executes in milliseconds, which is orders of magnitude faster than generating the native waveform.  }
    \label{fig:compression}
\end{figure*}

A key demonstration of this work is that this compression framework successfully extends to highly eccentric signals. Accurately capturing the complex phase evolution and inter-orbit harmonics does require significantly more interpolation nodes than is typically required for circular waveforms. However, the total storage footprint remains highly feasible, requiring on the order of a few hundred gigabytes to store the compressed representation of the entire 88.6 million template search space. To support a bank of this magnitude, we extended the existing \texttt{PyCBC} compression infrastructure. While we introduced a quadratic interpolation scheme that provides a modest reduction in the required number of nodes, the primary upgrade was a significantly accelerated, non-serial algorithm for automatically identifying the optimal nodal spacing. As illustrated in Fig.~\ref{fig:compression}, this identification algorithm is self-checking, iteratively populating nodes until the match between the interpolated template and the original signal exceeds a strict tolerance of 0.999. Rather than inserting a single node per pass, each iteration adds the midpoints of a batch of the worst-fitting inter-node segments, so the node set converges in a small number of passes over the waveform. Termination is decided on the global mismatch between the reconstructed and original waveform rather than on the per-segment error estimates, which can individually fall below the target while the full reconstruction does not. The full dynamic decompression executes in milliseconds per template, removing waveform generation as a computational bottleneck.

\section{Matched Filtering with Multibanding and Ratio-Filtering Details}
\label{app:multiband}

\begin{figure*}[t]
    \centering
    \includegraphics[width=\textwidth]{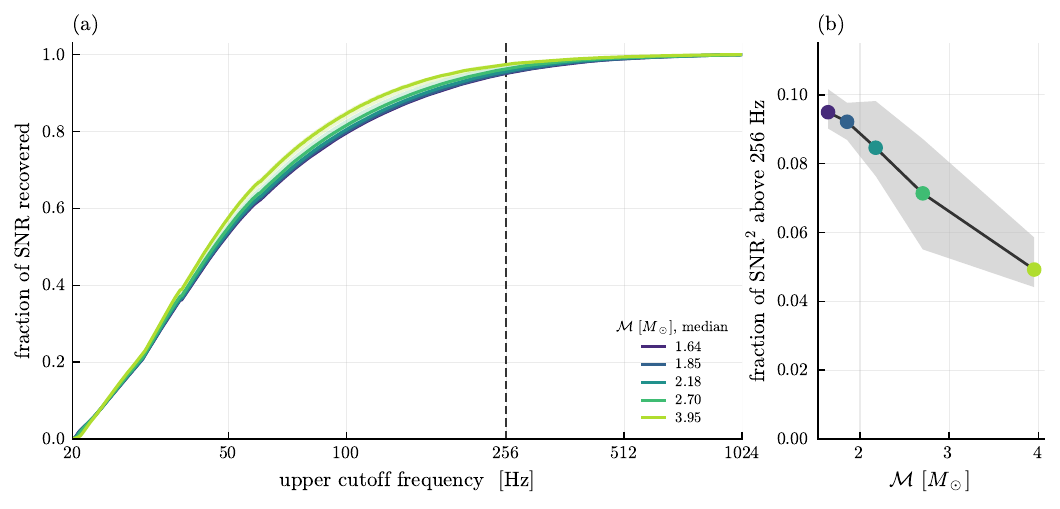}
    \caption{The accumulation of the signal-to-noise ratio across the frequency band for various NSBH chirp masses. Panel (a) plots the fraction of recovered SNR as a function of the upper cutoff frequency. A vertical dashed line marks the $256\text{ Hz}$ boundary. Panel (b) shows the fraction of squared SNR accumulated above $256\text{ Hz}$ as a function of the binary chirp mass. The grey shaded band indicates the variance across the targeted parameter space. The vast majority of the signal power is constrained to the low-frequency regime.}
    \label{fig:snr_accumulation}
\end{figure*}

\begin{figure*}[t]
    \centering
    \includegraphics[width=\textwidth]{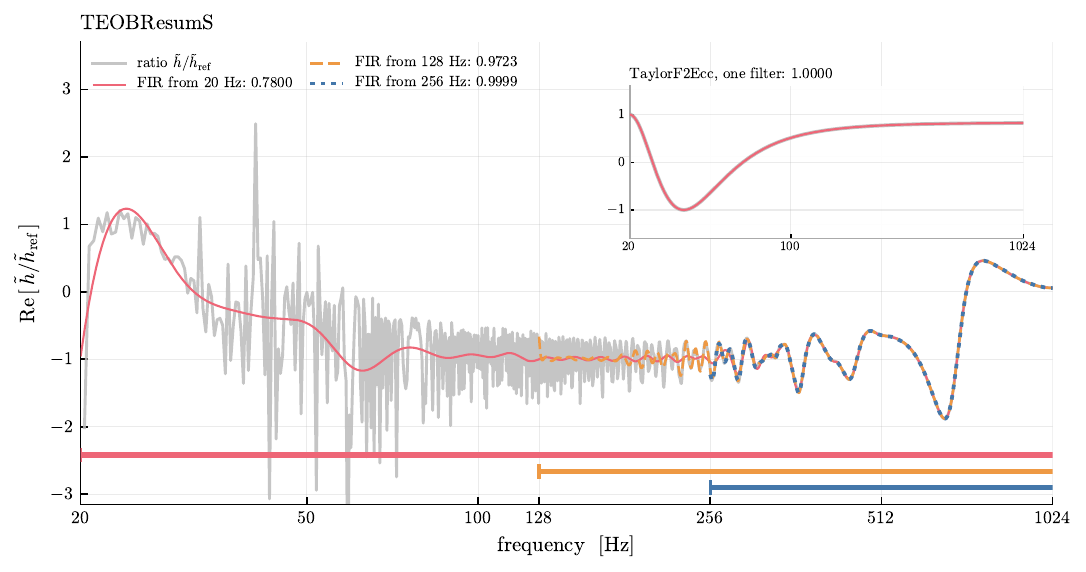}
    \caption{The efficacy of ratio filtering applied to highly eccentric signals. The main panel displays the frequency-domain ratio between an eccentric template and a reference waveform. At low frequencies, inter-orbit harmonics generate rapid oscillations in this ratio. The solid and dashed lines represent finite impulse response (FIR) filters initialized from different starting frequencies. Filters starting at $20\text{ Hz}$ fail to accurately capture the highly oscillatory low-frequency ratio. Filters constructed from $128\text{ Hz}$ and $256\text{ Hz}$ achieve matches near unity because the binary circularizes and harmonic amplitudes decay at high frequencies. The inset shows the smooth, non-oscillatory ratio curve for TaylorF2Ecc, whose eccentricity enters as secular dephasing only. Standard ratio-filtering would work for TaylorF2Ecc at the cost of the more accurate waveform modelling.}
    \label{fig:ratio_filter}
\end{figure*}

\begin{figure*}[t]
    \centering
    \includegraphics[width=\textwidth]{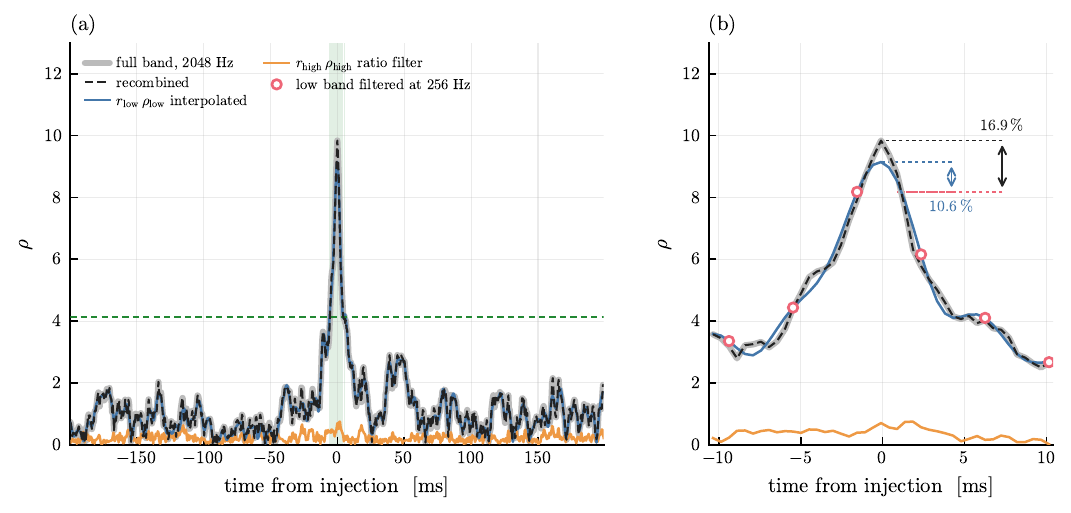}
    \caption{Validation of the hierarchical multibanded ratio-filtering approach for reconstructing the full-band SNR time series. Panel (a) shows the SNR traces surrounding an injected signal, including the full-band reference at $2048\text{ Hz}$, the coarsely sampled low-frequency band at $256\text{ Hz}$, the temporally reconstructed low-frequency band, and the high-frequency ratio filter contribution. Panel (b) provides a detailed view of the merger peak. The recombined SNR curve accurately reconstructs the full-band peak. Relying solely on the discretely sampled low-frequency band or omitting the high-frequency ratio filter contribution results in a measurable deficit in the recovered peak SNR. }
    \label{fig:snr_timeseries}
\end{figure*}

For long-duration, low-mass binaries, the majority of the signal-to-noise ratio (SNR) accumulates at low frequencies. Figure~\ref{fig:snr_accumulation} illustrates this fractional accumulation as a function of the upper cutoff frequency. This physical distribution motivates the use of a multibanded filtering architecture, an approach introduced to gravitational-wave searches by the MBTA pipeline~\cite{Adams:2015xza,Aubin:2020goo,Allene:2025saz} and also similar to the multi-rate filtering employed by the GstLAL pipeline~\cite{Cannon:2011vi,Sachdev:2019vvd}. By splitting the analysis, the early inspiral can be explicitly filtered at a severely reduced sample rate. However, naive multibanding still requires recovering the full sample rate. The floating-point operations required for this temporal reconstruction can largely offset the computational savings gained by evaluating the low-frequency band at a lower sample rate, though the reduced memory footprint does allow for more efficient utilization of the hardware memory hierarchy.

To achieve a substantial reduction in total floating-point operations, we must limit the high-frequency evaluation and reconstruction steps. We accomplish this by pairing multibanding with ratio-filtering in a hierarchical triggering scheme. Previous circular searches demonstrated that evaluating data against the slowly varying ratio between adjacent templates provides substantial speed improvements ($\sim 8\times$)~\cite{Nitz:2026fyr}. While this method cannot be applied across the full bandwidth of highly eccentric signals because inter-orbit harmonics prevent the low-frequency template ratios from dividing cleanly into a compact form (see Fig.~\ref{fig:ratio_filter}), the amplitude of these harmonics decays as the binary circularizes towards merger. The ratio-filtering approach therefore remains highly effective in the high-frequency regime, allowing for the construction of high-accuracy, short-duration filters for the late inspiral. Based on this, we split the frequency space into two bands: 20--256~Hz, and 256--1024~Hz, the upper edge being the Nyquist frequency of the 2048~Hz base sample rate. The low band is filtered as a complex series at the critical rate, a sample spacing of $1/256$~s, while the high-frequency contribution is evaluated with 251-tap FIR ratio filters designed to a minimal match of $0.999$ and achieving a median match of $0.99996$ across the bank. Data are analyzed in 512~s segments and the ratio filters are applied by overlap-save in blocks of 1024 samples. 

The data is first matched-filtered against the low-frequency templates at the reduced sample rate. To avoid missing a narrow SNR peak due to the coarse temporal spacing and missing high-frequency content, the high-frequency contribution is calculated and the low-frequency SNR is reconstructed at the full sample rate only when the coarsely sampled SNR exceeds an initial SNR threshold ($\sim 4$ here). The upsampling is done using a 128 tap Kaiser-windowed sinc function. This threshold accounts for both the missing SNR from the high-frequency part of the signal and the offset of the coarse samples from the true peak along with a statistical noise fluctuation. This was tuned on Gaussian noise to avoid
losing more than $\mathcal{O}(1\%)$ of triggers from a single template that would exceed the chosen target SNR threshold (5.5 here) after full reconstruction. Note that this threshold would need to be lowered for searches that require a lower single-detector threshold, such as high-mass BBH sources, and we would expect reduced computational savings in such cases from this hierarchy. If the initial threshold must be set low enough that nearly every template passes it, each would require both the full low-frequency reconstruction and the full high-frequency ratio filter, roughly halving the speedup. The interpolation cost can still largely be avoided by folding it into the trigger stage, and the high band remains ratio-filtered, so this is a conservative bound. 

Figure~\ref{fig:snr_timeseries} demonstrates that this hierarchical multibanded ratio-filtering approach accurately reconstructs the full-band SNR time series. The residual variance between the reconstructed SNR and the direct time-domain filtering SNR is strictly bounded by the known uncertainties derived from the waveform compression tolerance and the mismatch of the filter ratio. Because both pipelines filter the same compressed templates, the compression error is common to both and cancels in the difference, leaving the ratio-filter mismatch of $4\times10^{-5}$ to set the scale. Such an approximation injects an expected variance proportional to the square root of the effective mismatch~\cite{Lindblom:2008cm,Baird:2012cu}, here $\sigma(\Delta\rho)\simeq0.009$ against a measured $0.017$. The remainder has been tracked to increased numerical noise from the dynamic range of the ratio filters themselves and could be reduced with the tradeoff of longer filters. 

\section{Signal Consistency Implementation and Observational Validation}
\label{app:consistency}

To maximize the efficiency of the underlying Fourier transforms, templates are processed and filtered in batches of 128; the precise batch size is a tuning parameter rather than a property of the method. This batching allows the pipeline to leverage highly optimized, vectorized batched FFT operations, which is critical for maintaining high throughput for small FFT sizes. The trigger generation and subsequent signal consistency tests operate directly on these batched outputs. 

Matched-filter searches rely heavily on signal consistency tests, such as the Allen $\chi^2$ test~\cite{Allen:2004gu}, to distinguish true gravitational-wave signals from non-Gaussian transient noise (glitches). We adapt the standard $\chi^2$ test to accommodate our hierarchical multiband architecture. The traditional multi-bin $\chi^2$ statistic is calculated explicitly within the low-frequency band. Because the high-frequency contribution is evaluated dynamically via ratio-filtering and represents a smaller fraction of the total signal power, we treat the entire high-frequency band as a single additional $\chi^2$ bin. This is achieved by combining the independently calculated SNRs from both bands using their proper normalizations. As an added safeguard against high-frequency glitches, we enforce an additional consistency check requiring that the high-frequency band contribute less than $30\%$ of the total SNR.

\begin{figure*}[t]
    \centering
    \includegraphics[width=\textwidth]{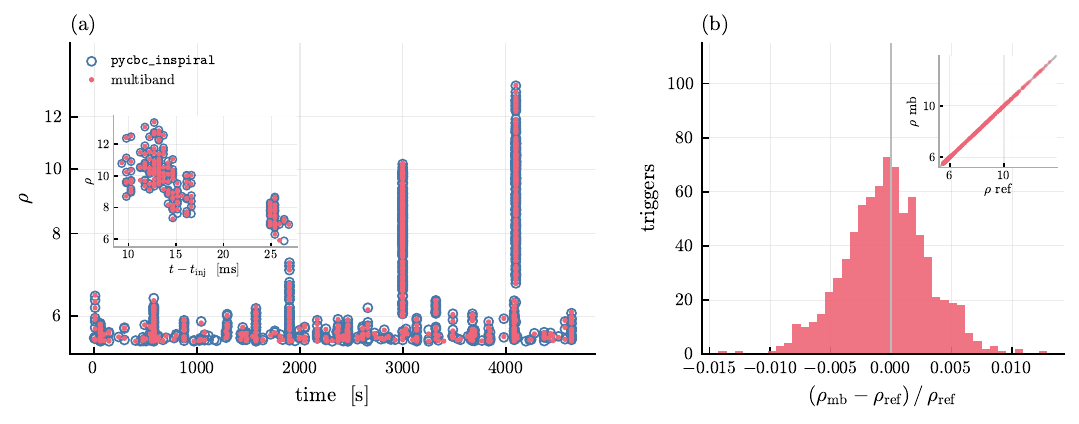}
    \caption{A comparison between the rapid multiband search methodology and the standard PyCBC Inspiral direct-filtering approach. Panel (a) overlays the recovered SNR of triggers over time for both pipelines. The inset zooms in on a cluster of triggers to demonstrate alignment. Panel (b) provides a histogram of the fractional difference in recovered SNR between the two methods. The distribution is tightly centered at zero, and the variance remains within the expected numerical bounds dictated by the compression and ratio-filtering approximations along with numerical error. Several triggers are missing in the multiband method due to differences in clustering and trigger selection, but do not reflect differences in internally constructed SNR timeseries. }
    \label{fig:pycbc_triggers}
\end{figure*}

\begin{figure*}[t]
    \centering
    \includegraphics[width=\textwidth]{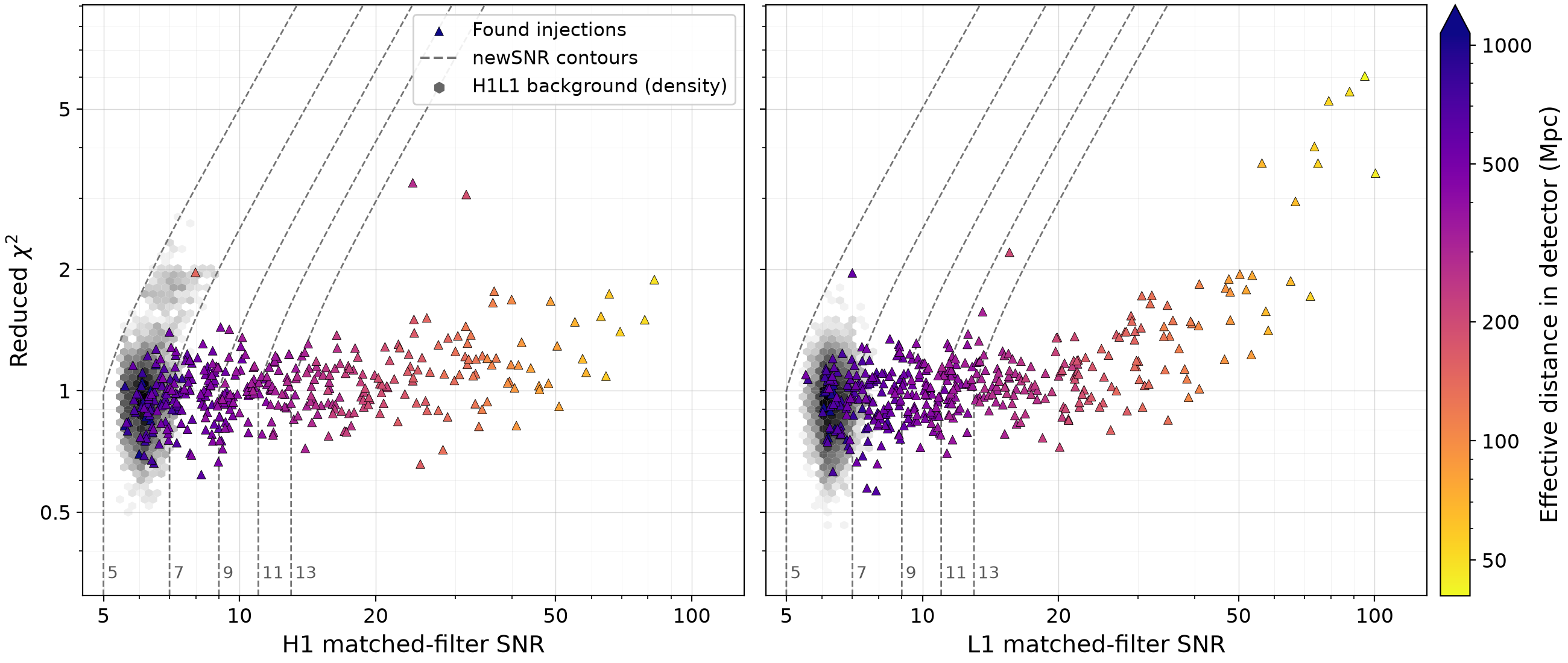}
    \caption{Performance of the adapted signal-consistency test in the multiband pipeline.
    The reduced $\chi^2$ is shown as a function of matched-filter SNR for
    single-detector triggers in H1 (left) and L1 (right) from one week of O4a data
    analyzed with the \texttt{TEOBResumS} bank. Grey shading shows the density of
    H1L1 background triggers. Triangles show found injections from the
    \texttt{TEOBResumS} with higher order modes and aligned spin campaign, coloured by their effective distance in each
    detector. Dashed lines are contours of constant reweighted SNR from 5 to 13.
    Background triggers are concentrated at low SNR with reduced $\chi^2 \sim 1$,
    while found injections maintain reduced $\chi^2 \sim 1$ up to SNR $\sim 50$, with a mild rise for the loudest injections,
    and separate cleanly from the background.}
    \label{fig:signal_consist}
\end{figure*}

To mitigate the cost of evaluating the signal consistency tests, we implement a descending-SNR heuristic. Instead of evaluating the $\chi^2$ statistic for every trigger, we identify the trigger with the highest preliminary SNR within a predefined time window ($1s$) and calculate its $\chi^2$ value first. This is used as the representative value for all triggers within its time window and for its batch of templates (e.g. not for the entire search space). The algorithm then proceeds to the next loudest trigger which does not already have an assigned value. This heuristic is effective for two reasons: first, the loudest trigger associated with a real astrophysical signal typically represents the best template match and yields the lowest $\chi^2$ value; second, the inverse is generally true for glitch-induced triggers, which often produce scattered, high-SNR triggers with poor $\chi^2$ fits. Because this assignment extends across the templates within a batch, a sub-dominant trigger may inherit the $\chi^2$ of a louder trigger from a neighboring template; any resulting effect on candidate ranking is present equally in the empirically measured background and is therefore accounted for when assigning significance.

To validate the accuracy and assess the computational performance of the new pipeline, we perform a direct comparison against the standard \texttt{PyCBC Inspiral} code which uses direct FFT-based matched filtering. 
Figure~\ref{fig:pycbc_triggers} demonstrates the trigger equivalence between the two pipelines using a controlled data set: 8192~s of Gaussian noise coloured by an O4 sensitivity curve, with no injections, filtered against a single sub-bank of 50\,393 templates. The inset of Figure~\ref{fig:pycbc_triggers} shows the zoomed-in view of the trigger alignment. The temporal alignment of the recovered triggers is highly consistent. When comparing the fractional difference in the recovered SNR of matching triggers, the distribution is tightly centered at zero. The observed variance is strictly within the expected numerical bounds dictated by the waveform compression tolerance, the ratio-filtering mismatch, and baseline numerical noise. A small number of triggers appear disjointed between the two pipelines; this is an expected artifact of differing clustering algorithms. This is specifically due to the use of fixed time bins in the multiband approach versus rolling windows in the standard pipeline, rather than disagreement in the internally constructed SNR time series.

Computational speed is only useful if the approximations it relies on preserve the ability to distinguish astrophysical signals from noise. This is particularly relevant for the adapted $\chi^2$ in the previous sections, which combines multi-bin low-frequency information with a single high-frequency bin and relies on the descending-SNR heuristic. Figure~\ref{fig:signal_consist} shows the reduced $\chi^2$ against matched-filter SNR for single-detector triggers from one week of O4a data. Background triggers from O4a data are concentrated at low SNR with reduced $\chi^2 \sim 1$, with a tail toward larger $\chi^2$ produced by non-Gaussian transient noise. Found injections instead remain at reduced $\chi^2 \sim 1$ up to SNR $\sim 50$ and separate clearly from the background in both detectors. The mild rise in $\chi^2$ for the loudest injections is expected from the residual mismatch between the $(2,2)$-mode templates and signals containing higher-order modes, whose contribution to $\chi^2$ grows with SNR. These signals are penalized only weakly by the reweighting and remain at high reweighted SNR. The adapted signal-consistency test therefore performs as intended on real detector data, suppressing background without rejecting the eccentric signals targeted by the search. 

\section{Single-Node Benchmarks and Contention Profiling}
\label{app:benchmarking}

\begin{figure*}[t]
    \centering
    \includegraphics[width=\textwidth]{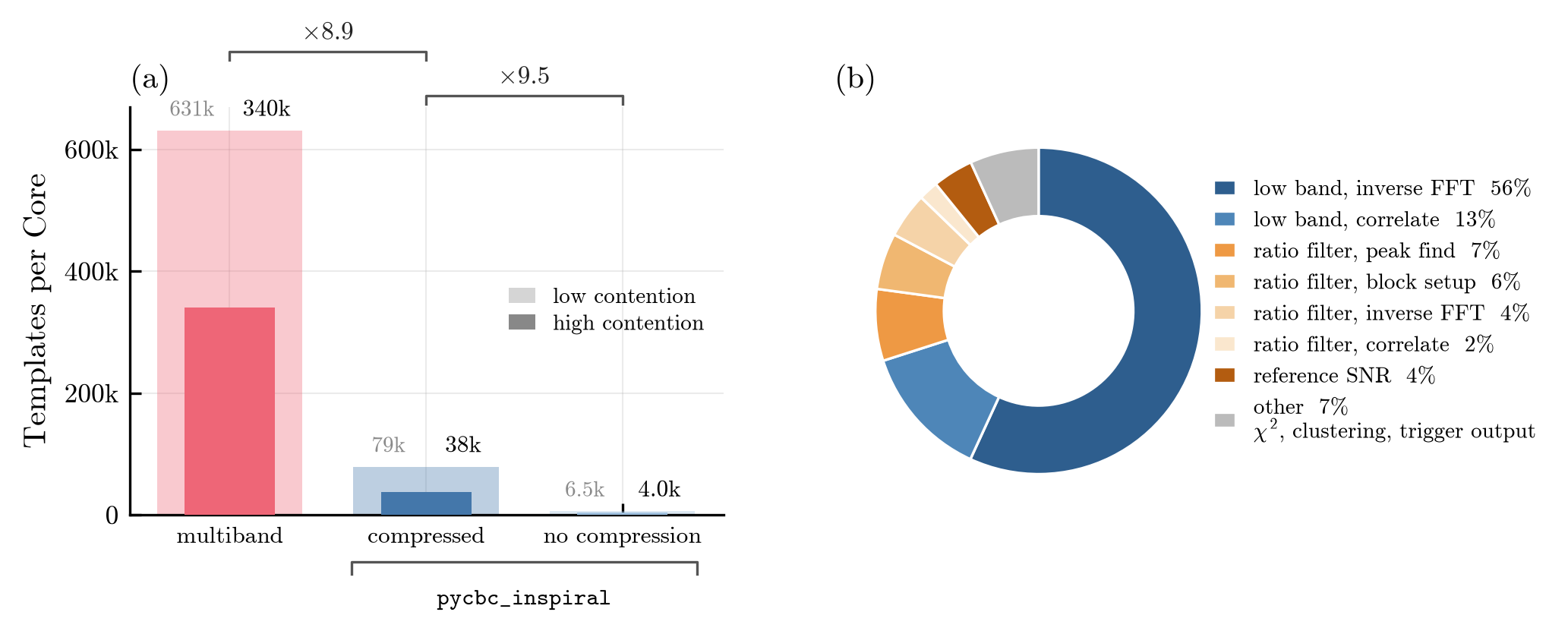}
    \caption{The computational throughput and cost breakdown of the rapid search pipeline run on a AMD AI Max+ 395 processor normalized to a single core performance. Results are shown both for the expected case of a fully loaded machine (high contention) and one where a single process runs (low contention). Panel (a) compares the throughput in templates per core for the new hierarchical multiband approach, standard filtering without waveform compression, and the baseline PyCBC Inspiral code that uses waveform compression. The new pipeline achieves an $8.9\times$ speed improvement over the compressed PyCBC Inspiral baseline on a fully loaded machine, and $8.0\times$ at low contention. Panel (b) details the remaining computational cost of the optimized matched filtering code. The total operational cost is strictly dominated by the batched inverse FFTs required for the low-band filtering, followed by the operations required for the ratio filtering. Only a small fraction comes from trigger handling and signal consistency tests.}
    \label{fig:pycbc_throughput}
\end{figure*}

The primary advantage of this hierarchical multiband architecture is its computational efficiency. We benchmarked the throughput of the search pipeline on an AMD AI Max+ 395 processor, evaluating both low-contention (single process) and high-contention (fully loaded machine) scenarios. As shown in Fig.~\ref{fig:pycbc_throughput}(a), the new pipeline achieves an $8.9\times$ speed improvement over the baseline \texttt{PyCBC Inspiral} code on a fully loaded machine, and $8.0\times$ when a single process has the machine to itself. This recovers, for eccentric templates, the speed improvement that ratio-filtering alone provides only for quasi-circular ones~\cite{Nitz:2026fyr}.

Figure~\ref{fig:pycbc_throughput}(b) provides a breakdown of the computational cost for the optimized multiband pipeline. After implementing the hierarchical trigger heuristic and ratio-filtering, the total operational cost is dominated by the batched inverse FFTs required to evaluate the low-frequency band. The operations required for high-frequency ratio filtering represent the second-largest cost, while the optimized trigger handling and signal consistency tests contribute only a small fraction to the overall computational footprint.

\section{Relative Sensitivity and Comparison with Simplified Waveform Models}
\label{app:relative_vt}

\begin{figure*}[t]
    \centering
    \includegraphics[width=\textwidth]{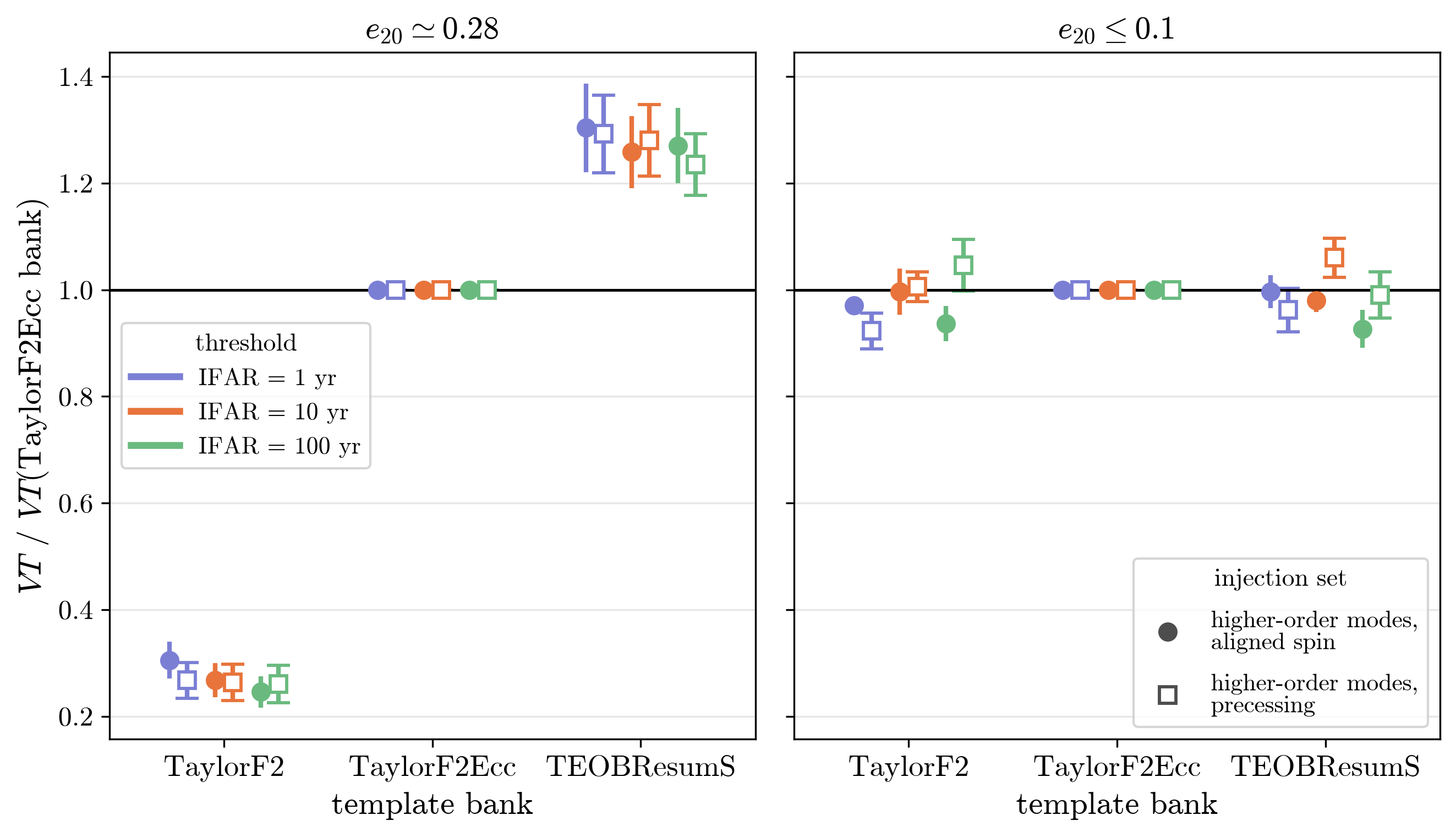}
    \caption{Sensitive volume-time $VT$ of each template bank relative to the \texttt{TaylorF2Ecc} bank, measured from prototype injection campaigns on one week of O4a H1 and L1 data. Injections are distributed uniformly in eccentricity at 20~Hz, over $0.20 \le e_{20} \le 0.28$ in the left panel and $0 \le e_{20} \le 0.1$ in the right panel. Results are shown for two injection sets generated with \texttt{TEOBResumS} including higher-order modes, with aligned spins (filled circles) and precessing spins (open squares), at inverse false-alarm-rate thresholds of 1, 10, and 100 years. Error bars denote Monte Carlo errors of the ratio. For highly eccentric signals, the \texttt{TEOBResumS} bank recovers
    $\sim 25$--$30\%$ more sensitive volume than the \texttt{TaylorF2Ecc} bank,
    while the \texttt{TaylorF2} bank reaches only $\sim 25$--$30\%$ of the
    \texttt{TaylorF2Ecc} sensitivity. At low eccentricity, all three banks are consistent within uncertainties.}
    \label{fig:relative_sensitivity}
\end{figure*}

To quantify the direct sensitivity benefit of utilizing full-physics EOB waveforms over simplified models, Figure~\ref{fig:relative_sensitivity} compares the sensitive volume-time $VT$ of each bank with the \texttt{TaylorF2Ecc} bank on a representative prototype injection campaign across one week of O4a data. Across both injection sets and all inverse false-alarm-rate (IFAR) thresholds, the \texttt{TEOBResumS} bank increases the sensitive volume by $\sim 25$--$30\%$ relative to \texttt{TaylorF2Ecc}.

The improvement is achieved even though the \texttt{TEOBResumS} bank is roughly seventy times larger than the \texttt{TaylorF2Ecc} bank and therefore accumulates a correspondingly larger background of noise triggers. The signal-to-noise ratio recovered by modeling inter-orbit harmonics and apsidal precession more than compensates for this additional trials factor. In contrast, the \texttt{TaylorF2} bank retains only $\sim 25$--$30\%$ of the \texttt{TaylorF2Ecc} sensitive volume, roughly a fifth of that achieved by the \texttt{TEOBResumS} bank. This confirms that searches neglecting eccentricity are largely blind to this population. For the low-eccentricity population, all three banks recover the same sensitive volume within statistical uncertainties. This is expected, since \texttt{TaylorF2Ecc} is already accurate below $e_{20} \sim 0.1$. In both populations, the results are similar for aligned and precessing injections, indicating that unmodeled spin precession does not erase the benefit of the more accurate eccentric waveform model.

\end{document}